\documentclass[aps,showkeys,preprintnumbers,amsmath,epsf,amssymb,epsfig,nofootinbib]{revtex4}
\input epsf.sty
\usepackage{bm}
\usepackage{graphicx}
\draft

\newcommand {\de}{\delta}

\newcommand {\La}{\Lambda}

\newcommand {\al}{\alpha}

\newcommand {\na}{\nabla}
\newcommand {\fr}{\frac}
\newcommand {\vphi}{\varphi}
\newcommand {\ca}{{\cal A}}

\newcommand {\beg}{\begin{equation}}
\newcommand {\en}{\end{equation}}
\newcommand {\bega}{\begin{eqnarray}}
\newcommand {\ena}{\end{eqnarray}}
\begin{document}
\title{Power spectrum of large-scale magnetic fields from
Gravitoelectromagnetic inflation with a decaying cosmological
parameter}
\author{ $^{1,2}$ Federico Agust\'{\i}n Membiela \footnote{
E-mail address:membiela@argentina.com}, $^{1,2}$ Mauricio Bellini
\footnote{E-mail address: mbellini@mdp.edu.ar}}
\address{$^{1}$ Departamento de F\'{\i}sica, Facultad de Ciencias Exactas y
Naturales, Universidad Nacional de Mar del Plata, Funes 3350,
(7600) Mar del Plata,
Argentina.\\
$^{2}$ Consejo Nacional de Investigaciones Cient\'{\i}ficas y
T\'ecnicas (CONICET). }

\begin{abstract}
Introducing a variable cosmological parameter $\Lambda (t)$ in a
geometrical manner from a 5D Riemann-flat metric, we investigate
the origin and evolution of primordial magnetic fields in the
early universe, when the expansion is governed by a cosmological
parameter $\Lambda (t)$ that decreases with time. Using the
gravitoelectromagnetic inflationary formalism, but without the
Feynman gauge, we obtain the power of spectrums for large-scale
magnetic fields and the inflaton field fluctuations during
inflation. A very important fact is that our formalism is {\em
naturally non-conformally invariant}.
\end{abstract}

\maketitle

\section{Introduction}

The origin of the primordial magnetic fields has been subject of a
great amount of research\cite{mag}. The existence, strength and
structure of these fields in the intergalactic plane, within the
Local Superclusted, has been scrutinized recently\cite{ocho}. Many
spiral galaxies are endowed with coherent magnetic fields of $\mu
G$ (micro Gauss) strength~\cite{1,2,3,4,5,6}, having approximately
the same energy density as the cosmic microwave background
radiation (CMBR). In particular, the field strength of our galaxy
is  $B \simeq 3 \times 10^{-6} \  G$, similar to that detected in
high redshift galaxies~\cite{uno} and damped Lyman alpha
clouds~\cite{dos}. Limits imposed by the high isotropy of CMB
photons, obtained from the COBE data\cite{clarkson} restrict the
present day strength of magnetic fields on cosmological scales to
$10^{-9}\,G$. It is very mysterious that magnetic fields in
clusters of galaxies [i.e., on scales  $ \sim \, {\rm Mpc}$], to
be coherent. The origin of these magnetic fields is not well
understood yet. The seeds of these fields could be in the early
inflationary expansion of the universe, when these fields were
originated. Therefore, the study of its origin and evolution in
this epoch should be very important to make predictions in
cosmology\cite{giovannini}. During inflation the extension of the
causally connected regions grows as the scale factor and hence
faster than in the decelerated phase. This solves the horizon
problem. Furthermore, during inflation the contribution of the
spatial curvature becomes very small. The way inflation solves the
curvature problem is by producing a very tiny spatial curvature at
the onset of the radiation epoch taking place right after
inflation. The spatial curvature can well grow during the
decelerated phase of expansion but it will be always subleading
provided inflation lasted for sufficiently long time. It is
natural to look for the possibility of generating such a
large-scale magnetic field during inflation. However, the FRW
universe is conformal flat and the Maxwell theory is conformal
invariant. Therefore, the conformal invariance must be broken to
generate non-trivial magnetic fields. Various conformal symmetry
breaking mechanisms have been proposed so far\cite{varios}.

Gravitoelectromagnetic inflation was developed very recently with
the aim to describe, in an unified manner, the inflaton,
gravitatory and electromagnetic fields during
inflation\cite{g1,g2}. In this formalism all the 4D sources have a
geometrical origin. This formalism can explain the origin of seed
magnetic fields on cosmological scales observed today.
Gravitoelectromagnetic inflation was constructed from a 5D vacuum
state on a $R^{A}\,_{BCD}=0$ globally flat metric. As in all Space
Time Matter (STM) models\cite{.}, the 4D sources are geometrically
induced when we take a foliation on the fifth coordinate which is
spacelike and noncompact. However, in the previous works was used
the Feynman gauge in order to simplify the structure of the field
equations.

In this letter we shall extend this formalism without using the
Feynman gauge. As we shall see, the field equations become
coupled, which has interesting physical consequences. We shall
study the origin and evolution of the seed magnetic fields in a
$\Lambda(t)$ (with $\dot\Lambda <0$) dominated early universe,
from a 5D vacuum state. Finally, we shall try to explain why
(since have been observed) the large-scale magnetic fields are
coherent. \\

\section{Vector fields in 5D vacuum}

We begin considering a 5D manifold $\cal{M}$ described by a
symmetric metric $g_{AB}=g_{BA}$\footnote{In our conventions
capital Latin indices run from $0$ to $4$, greek indices run from
$0$ to $3$ and latin indices run from $1$ to $3$.}. This manifold
$\cal M$ is mapped by coordinates $\{x^A\}$.
 \beg
    dS^2=g_{AB}dx^A dx^B,\label{meta}
 \en
 which, we shall consider as Riemann-flat $R^A_{BCD}=0$. To introduce
 the fields we can define an action in $\cal M$.
 \beg
    \mathcal{S}=\int d^5x\sqrt{-g}\left[\fr{^{(5)}\,R}{16\pi
    G}+\mathcal{L}_{(\ca_B,\na_B\ca_{C})}\right],
 \en
where $\ca_B=(A_\mu,\vphi)$ and $\ca^B=(A^\mu,-\vphi)$ are
respectively the covariant and contravariant 5-vector potentials,
$^{(5)}\,R$ is the 5D scalar curvature and $\na_B$ denotes de
covariant derivative. We shall consider these fields as minimally
coupled to gravity. In this space the fields will be free of
potential energy, so the lagrangian density is of the form

 \beg
    \mathcal{L}=-\fr{1}{4}Q_{BC}Q^{BC} = -\frac{1}{4} F_{BC} F^{BC} - \frac{5\gamma^2}{4} \left(
    \nabla_D A^D\right)^2, \label{lag1}
 \en
with $Q_{BC}\doteq F_{BC}+\gamma \,g_{BC}(\na_D
{\ca^D})^2$\cite{g1,g2}. Notice the importance of the last term in
$Q_{BC}$, which has been included to add the symmetries of gravity
through $g_{AB}$. This term is the responsible for the rupture of
conformal invariance in the Lagrangian (\ref{lag1}). The Faraday
tensor is antisymmetric $F_{BC}=\na_B \ca_C-\na_C \ca_B$, and the
last term in (\ref{lag1}) is a 5D ``gauge-fixing'' term. The
Lagrange equations are then

 \beg \label{mov}
    \na^F\na_F \ca^B-{R^B}_F \ca^F-\al\na^B\na_F \ca^F=0,
 \en
where $\al=1-\fr{5}{2}\gamma^2$. We shall consider the following
non commutative algebra for $A^C$ and $\bar\Pi^B = {\partial {\cal
L}\over \partial\left(\nabla_0 A_B\right)}=F^{B0}-g^{B0}\nabla_C
A^C$
\begin{eqnarray}
&& \left[A^C(t,\vec r,\psi),\bar\Pi^B(t,\vec{r}',\psi)\right]
=i\,g^{CB} g^{tt} \left|\frac{^{(5)} g_0}{^{(5)}g}\right|
\,\delta^{(3)}(\vec r - \vec{r}'), \\
&& \left[A^C(t,\vec r,\psi),A^B(t,\vec{r}',\psi)\right] =
\left[\bar\Pi^C(t,\vec
r,\psi),\bar\Pi^B(t,\vec{r}',\psi)\right]=0.
\end{eqnarray}
Here, $\bar\Pi^t= - g^{tt}\left(\nabla_C A^C\right)$ and $\left|
{^{(5)} g_0 \over ^{(5)} g}\right|$ is the inverse of the
normalized volume of the manifold (\ref{meta}).

In this work we shall choose the generalized Lorentz gauge, so
that the field equations in (\ref{mov}) will become

\begin{eqnarray}
&& \na_F A^F =0,  \label{mov1} \\
&&  \na^F\na_F \ca^B-{R^B}_F \ca^F=0. \label{mov2}
\end{eqnarray}
The equation (\ref{mov1}) describes the conservation of the
vectorial field ${\cal A}[A^F(x^B)]$ on the 5D Riemann-flat metric
we shall consider in this work. The equation (\ref{mov2}),
describes the motion of the components $A^F$ on a metric with
Ricci-tensor components: $R_{SL}$. In our case will be null on the
5D flat metric we shall use.

The observers are constrained to see only an hypersurface of the
manifold. We can define an hypersurface $\Sigma_f$ using a
function $f=f(x^A)$ and taking the constrain equation
$f(x^A)=cte$. The hypersurface is then mapped by coordinates
$\{y^\nu\}$ with ($\nu=0,1,2,3$) and has an induced metric
$g_{\mu\nu}$. The direction normal to $\Sigma_f$ is $n_A$ defined
by $\na_A f$.

The 5D quantities in $\cal M$ have their counterparts in the 4D
hypersurface $\Sigma_F$. Physical quantities in the brane are
defined by identifying the parallel parts to the hypersurface and
the normal ones. On the other hand, geometrical quantities in the
brane are constructed from the induced metric $g_{\mu\nu}$.\\

\subsection{The 5D Riemann-flat metric with decaying parameter}

In particular, in this letter we are interested to deal with the
following Riemann-flat metric\cite{coscos}

 \beg
    dS^2=\psi^2\fr{\La(t)}{3}dt^2-\psi^2
    e^{2\int^t_0 d\tau\sqrt{\La(\tau)/3}}dr^2-d\psi^2, \label{met1}
 \en
where $dr^2=dx^i\de_{ij}dx^j$ is the euclidean line element in
cartesian coordinates and $\psi$ is the space-like extra
dimension. Adopting natural units ($\hbar=c=1$) the cosmological
parameter $\La(t)$ (with $\dot\Lambda <0$), has units of
$(length)^{-2}$. The metric (\ref{met1}) is very interesting to
study the evolution of the gravitoelectromagnetic (vectorial)
field, because is Riemann-flat, but has some conections
$\Gamma^{C}_{\,DE}\neq 0$. This fact is very important when we
consider the covariant derivative of $A^F$.

The equations of motion for the components of the vectorial field
${\cal A}$ are
\begin{eqnarray}
&& \frac{\partial^2 A_4}{\partial t^2} + \left[
3\sqrt{\frac{\Lambda}{3}}-\frac{\dot\Lambda}{2\Lambda}\right]
\frac{\partial A_4}{\partial t} - \frac{\Lambda}{3} e^{-2\int
\sqrt{\frac{\Lambda}{3}} dt}\, \nabla^2 A_4 - \frac{\Lambda}{3}
\left[ \psi^2 \frac{\partial^2 A_4}{\partial\psi^2}+ 6\psi
\frac{\partial A_4}{\partial \psi} + 4
A_4 \right] =0,  \label{a1} \\
&& \frac{\partial^2 A_0}{\partial t^2} +
\left[5\sqrt{\frac{\Lambda}{3}}\right] \frac{\partial
A_0}{\partial t} - \frac{\Lambda}{3} e^{-2\int
\sqrt{\frac{\Lambda}{3}} dt} \nabla^2 A_0 + \left[
\left(\frac{\dot\Lambda}{\Lambda}\right)^2 - 5
\sqrt{\frac{\Lambda}{3}} \frac{\dot\Lambda}{2} -
\frac{\ddot\Lambda}{2\Lambda} + \left(2-\frac{\psi^2 {\cal
R}^0_{\,\,0}}{3}\right)
\Lambda\right] A_0 \nonumber  \\
&&- \frac{\Lambda}{3} \left[\psi^2\frac{\partial
A_0}{\partial\psi^2} + 2\psi \frac{\partial
A_0}{\partial\psi}\right] = \frac{\Lambda}{3} \left[2\psi
\frac{\partial A_4}{\partial t} + 2\psi^2 \sqrt{\frac{\Lambda}{3}}
\left( \frac{\partial A_4}{\partial\psi} + \frac{4}{\psi} A_4
\right)
\right] , \label{a2} \\
&& \frac{3}{\psi^2 \Lambda} \frac{\partial^2 A_i}{\partial t^2} +
\frac{3}{\psi^2\Lambda} \left[
3\sqrt{\frac{\Lambda}{3}}-\frac{\dot\Lambda}{2\Lambda}\right]
\frac{\partial A_i}{\partial t} - \frac{1}{\psi^2} e^{-2\int
\sqrt{\frac{\Lambda}{3}} dt} \nabla^2 A_i - \left[\frac{\partial
^2 A_i}{\partial \psi^2} + \frac{2}{\psi} \frac{\partial
A_i}{\partial\psi} + {\cal R}^j_{\,\,i} A_j\right]\nonumber \\
&& = \frac{2}{\psi^2} \left[\left.\vec\nabla\right|_{i} A_4 -
\sqrt{\frac{3}{\Lambda}} \left.\vec\nabla\right|_{i} A_0\right].
\label{a3}
\end{eqnarray}
Notice that the equation of motion (\ref{a3}) for $A_i$, depends
on the inflaton field $A_4\equiv \varphi$ and the electric
potential $A_0$. Furthermore, the equation of motion for the
electric potential $A_0$ depends on $\varphi$, which acts as a
source in (\ref{a2}). However, from the eq. (\ref{a1}) we see that
$\varphi$ is do not depends on the other fields. Furthermore, if
we consider the particular gauge $A_0(t,\vec r,\psi)=0$, we obtain
the following relevant equations:
\begin{eqnarray}
&& \frac{\partial^2 A_4}{\partial t^2} + \left[
3\sqrt{\frac{\Lambda}{3}}-\frac{\dot\Lambda}{2\Lambda}\right]
\frac{\partial A_4}{\partial t} - \frac{\Lambda}{3} e^{-2\int
\sqrt{\frac{\Lambda}{3}} dt}\, \nabla^2 A_4 - \frac{\Lambda}{3}
\left[ \psi^2 \frac{\partial^2 A_4}{\partial\psi^2}+ 6\psi
\frac{\partial A_4}{\partial \psi} + 4
A_4 \right] =0,  \label{aa1} \\
&&  \frac{\Lambda}{3} \left[2\psi \frac{\partial A_4}{\partial t}
+ 2\psi^2 \sqrt{\frac{\Lambda}{3}} \left( \frac{\partial
A_4}{\partial\psi} + \frac{4}{\psi} A_4\right)\right]=0 ,   \label{aa2} \\
&& \frac{3}{\psi^2 \Lambda} \frac{\partial^2 A_i}{\partial t^2} +
\frac{3}{\psi^2\Lambda} \left[
3\sqrt{\frac{\Lambda}{3}}-\frac{\dot\Lambda}{2\Lambda}\right]
\frac{\partial A_i}{\partial t} - \frac{1}{\psi^2} e^{-2\int
\sqrt{\frac{\Lambda}{3}} dt} \nabla^2 A_i - \left[\frac{\partial
^2 A_i}{\partial \psi^2} + \frac{2}{\psi} \frac{\partial
A_i}{\partial\psi} + {\cal R}^j_{\,\,i} A_j\right]\nonumber \\
&& = \frac{2}{\psi^2} \left.\vec\nabla\right|_{i} A_4 .
\label{aa3}
\end{eqnarray}
These are our equations of motion on the metric (\ref{met1}), once
we consider the generalized Lorentz gauge $\nabla_F A^F =0$, with
$A^0=0$.\\

\section{Effective 4D dynamics with a static foliation
$\psi=\psi_0$}

In order to study the effective evolution of the 4D universe, we
shall consider the metric (\ref{met1}) on the hypersurface $\psi =
\psi_0$. From the point of view of a relativistic observer, we are
saying that the pentavelocity $U^4 = {d\psi\over dS}\equiv
U^{\psi} =0$, such that $g_{CD} U^C U^D=1$. Furthermore, using the
changes of variables $\psi^2_0 = {3\over \Lambda_0}$, we obtain
the following effective 4D metric
\begin{equation}\label{met2}
\left.dS^2\right|_{eff} = \frac{\Lambda(t)}{\Lambda_0} dt^2 -
\frac{3}{\Lambda_0} e^{-\int^t_0 \sqrt{\frac{\Lambda}{3}} d\tau}
\, dr^2,
\end{equation}
where $\Lambda_0 = \Lambda(t=t_0)$ is some constant of $\Lambda$
at the initial time $t=t_0$ (which can be the Planckian time). If
now we require that the components $A_i$ describe photons on the
metric (\ref{met2}), we obtain $^{(4)}\Box A_i-{\cal R}^j_i\, A_j
=0$, where $^{(4)}\Box$ denotes the D'Alambertian operator on the
metric (\ref{met2}). The resulting equations on the 4D
hypersurface, are
\begin{eqnarray}
&& \left.\frac{\partial^2 A_4}{\partial t^2} + \left[
3\sqrt{\frac{\Lambda}{3}}-\frac{\dot\Lambda}{2\Lambda}\right]
\frac{\partial A_4}{\partial t} - \frac{\Lambda}{3} e^{-2\int
\sqrt{\frac{\Lambda}{3}} dt}\, \nabla^2 A_4 - \frac{\Lambda}{3}
\left[ \psi^2 \frac{\partial^2 A_4}{\partial\psi^2}+ 6\psi
\frac{\partial A_4}{\partial \psi} + 4 A_4
\right]\right|_{\psi=\psi_0} =0,  \label{aaa1} \\
&& \frac{\partial^2 A_i}{\partial t^2} + \left[
3\sqrt{\frac{\Lambda}{3}}-\frac{\dot\Lambda}{2\Lambda}\right]
\frac{\partial A_i}{\partial t} - \frac{\Lambda}{3} e^{-2\int
\sqrt{\frac{\Lambda}{3}} dt} \nabla^2 A_i -{\cal R}^j_{\,\,i}\, A_{\,\,j}=0, \label{aaa2} \\
&&  -\psi^2\frac{\Lambda}{3}\left\{\left.\left[ \frac{\partial ^2
A_i}{\partial \psi^2} + \frac{2}{\psi} \frac{\partial
A_i}{\partial\psi} \right] -\frac{2}{\psi^2}
\left.\vec\nabla\right|_{i} A_4\right\}\right|_{\psi=\psi_0} =
\left.\frac{\Lambda}{3} \left\{2\psi \frac{\partial A_4}{\partial
t} + 2\psi^2 \sqrt{\frac{\Lambda}{3}} \left(\frac{\partial
A_4}{\partial \psi} + \frac{4}{\psi}
A_4\right)\right\}\right|_{\psi=\psi_0}=0 , \label{aaa3}
\end{eqnarray}
where ${\cal R}^i_{\,\,i}=3/\psi^2_0$ and (\ref{aaa3}) play the
role of ligadure equations such that their solutions must be well
defined  $\forall\,\,\psi$. Notice that this equations are
arbitrary and were introduced to obtain an equation of motion for
photons in (\ref{aaa2}). The components
$\left.A^B\right|_{\psi=\psi_0}$, on the 4D hypersurface can be
written as
\begin{eqnarray}
&& \varphi(t,\vec r,\psi_0) =\frac{1}{(2\pi)^{3/2}} {\Large\int}
d^3k \left[\xi_k(t,\psi_0) e^{i\vec k .\vec r} + \xi^*_k(t,\psi_0)
e^{-i\vec k.\vec r}\right], \label{f1}
\\
&& A_{\mu} (t,\vec r, \psi_0) =\frac{1}{(2\pi)^{3/2}} {\Large\int}
d^3k  \sum_{\sigma=1,2} \left[\epsilon_{\mu}(\vec
k,\sigma)\,\xi^{(\sigma)}_k(t,\psi_0) e^{i\vec k .\vec r} +
\epsilon^*_{\mu} (\vec
k,\sigma)\,\left(\xi^{(\sigma)}_k(t,\psi_0)\right)^* e^{-i\vec
k.\vec r}\right], \label{f2}
\end{eqnarray}
where the two ``physical'' polarization 4-vectors satisfy
${\rm\epsilon}^{(\sigma)}.{\rm\epsilon}^{(\sigma')}=\delta^{\sigma\sigma'}$.
Furthermore, we shall define the modes of the fields as
\begin{eqnarray}
&& \xi^{(\sigma)}_k(t,\psi_0) = b^{(\sigma)}_{k}\,
\Psi_{k}(t), \label{mod1} \\
&& \xi_k(t,\psi_0)=a_{k}\, \chi_{k}(t), \label{mod2}
\end{eqnarray}
where the annhilation $\left(a_k, b^{(\sigma)}_k\right)$ and
creation $\left(a^{\dagger}_k,
\left(b^{(\sigma)}_k\right)^{\dagger}\right)$ operators comply
with the algebra
\begin{eqnarray}
&&\left[ b^{(\sigma)}_{\vec{k}}, \left( b^{(\sigma')}_{\vec{k'}}
\right)^{\dagger} \right] = \delta^{\sigma \sigma '} \,
\delta^{(3)} \left(\vec{ k} - \vec {k'}\right), \qquad
\left[\left( b^{(\sigma)}_{\vec{ k}}\right)^{\dagger}, \left(
b^{(\sigma ')}_{\vec{k'}}\right)^{\dagger} \right]= 0,
\label{co2} \\
&&\left[ a_{\vec{k}}, a^{\dagger}_{\vec{k'}} \right]=\delta^{(3)}
\left(\vec{ k} - \vec {k'}\right), \qquad \left[a^{\dagger}_{\vec{
k}}, a^{\dagger}_{\vec{k'}}\right] = 0. \label{co4}
\end{eqnarray}
Notice that we are considering $A_{\mu}(t,\vec r,\psi_0)$ as a
$U(1)$ gauge field on the effective 4D metric (\ref{met2}).\\

Finally, it is very interesting to study the magnetic field
fluctuations on the infrared (IR) sector (i.e., on scales very
bigger than the horizon radius) in the physical frame:
$\left.\left< B^2_{(phys)} \right>\right|_{IR} = \left.\left< B^i
B_i \right>\right|_{IR}=\left. \left< \left(\epsilon^{ijk}
 \nabla_j  A_k\right) \left(\epsilon_{ilm} \nabla^l A^m \right)\right>\right|_{IR}$,
which are related to the expectation value of the magnetic field
energy density $\left<\rho_B\right> = {1\over 8\pi}\left.\left<
B^i B_i \right>\right|_{IR}$

\section{Examples}

In order to ilustrate the formalism we can study two examples,
which are interesting for the cosmological expansion of the early
universe.

\subsection{de Sitter expansion: $\Lambda = \Lambda_0$}

As a first example we shall consider the case where $\Lambda =
\Lambda_0$. When we make $\psi_0=\sqrt{{3\over\Lambda_0}}$, this
case give us a de Sitter inflationary expansion of the universe
with tetra-velocities: $u^{\alpha} = (1,0,0,0)$ for a comoving
frame. Furthermore, the effective 4D line element is
\begin{equation}\label{sitter}
ds^2 = dt^2 - \frac{3}{\Lambda_0} \,e^{2 \sqrt{\Lambda_0/3} t} d{r
}^2,
\end{equation}
In this case the general solutions for the modes of photons
$\bar\Psi_{k}(t) = e^{{3\over 2} \sqrt{{\Lambda_0\over 3}}
t}\,\,\Psi_{k}(t)$ and the inflaton $\bar\chi_{k}(t)=e^{{3\over 2}
\sqrt{{\Lambda_0\over 3}} t}\,\,\chi_{k}(t)$, once normalized by
the conditions:
\begin{eqnarray}
&& \bar\Psi_{k} \left(\dot{\bar\Psi}_{k}\right)^*
-\left(\bar\Psi_{k}\right)^* \dot{\bar\Psi}_{k} =i,\\
&& \bar\chi_{k} \left(\dot{\bar\chi}_{k}\right)^* -
\left(\bar\chi_{k}\right)^* \dot{\bar\chi}_{k}
=i\,\sqrt{\frac{\Lambda_0}{3}}.
\end{eqnarray}

result to be
\begin{eqnarray}
&& \bar\Psi_{k}(t)= \frac{1}{2}
\sqrt{\frac{\pi}{\sqrt{\Lambda_0/3}\, m_0}} \,\,{\cal
H}^{(2)}_{\frac{\sqrt{9-4m^2_0}}{2}}\left[
k\,e^{-\sqrt{\frac{\Lambda_0}{3}} t}\right], \\
&& \bar\chi_{k}(t) = \frac{i}{2} \sqrt{\pi} \,{\cal
H}^{(2)}_{\frac{\sqrt{21}}{2}}\left[k
\,e^{-\sqrt{\frac{\Lambda_0}{3}} t}\right],
\end{eqnarray}
where $0 < m_0 < 3/2$ is some value of $m$. An estimation of the
squared magnetic field fluctuations on the infrared sector in the
physical and comoving frames, respectively, give us
\begin{eqnarray} && \left.\left< B^2_{(phys)}\right>\right|_{IR}
\simeq N\,\,\left(\frac{\Lambda_0}{3}\right)^2
\epsilon^{5-\sqrt{21}}\,\,e^{-2\sqrt{\frac{\Lambda_0}{3}} t}, \\
&& \left.\left< B^2_{(com)}\right>\right|_{IR} \simeq
N\,\,\left(\frac{\Lambda_0}{3}\right)^2
\epsilon^{5-\sqrt{21}}\,\,e^{2\sqrt{\frac{\Lambda_0}{3}} t}.
\end{eqnarray}
where $N$ and $\epsilon \ll 1$ are constants. The power spectrums
of $\left<\varphi^2\right>$ and $\left<B^2\right>$, go as
\begin{eqnarray}
&& {\cal P}^{\left<\varphi^2\right>}(k) \sim k^{3\left[1-\sqrt{1-\frac{4}{9}m^2_0}\right]}, \label{s1} \\
&& {\cal P}^{\left<B^2\right>}(k) \sim k^{5-\sqrt{21}} \sim
k^{0.42}. \label{s2}
\end{eqnarray}
Note that ${\cal P}^{\left<\varphi^2\right>}(k)$ becomes nearly
scale invariant for $m^2_0\ll 1$. Furthermore, we can relate the
parameter $m_0$ with the mass of the inflaton field ${\cal M}$ and
the Hubble parameter obtained in standard 4D theories of
inflation: $m^2_0 = { {\cal M}^2\over  H^2_0} \equiv {3{\cal M}^2
\over \Lambda_0} \ll 1$. On the other hand we see in (\ref{s2})
that the spectrum of $\left.\left<B^2\right>\right|_{IR}$ (and
hence the expectation value for the energy density due to magnetic
fields) on the infrared sector: $\left<\rho_B\right>$), go as
$\sim k^{0.42}$. This implies that magnetic fields should be more
intense on smaller scales. This agrees with observation, because
the observed strength of magnetic fields on galactic scales are
bigger than
whole of cosmological scales. \\

\subsection{Decaying cosmological parameter: $\Lambda=3p^2/t^2$}

A more interesting example can be obtainded making
$\Lambda=3p^2/t^2$ and $\psi_0=\sqrt{{3\over\Lambda_0}}$, where
$\Lambda$ is the cosmological parameter when inflation starts. In
this case the effective 4D line element is given by (\ref{met2}).
Since we require that $g_{CD} u^C u^D =1$, on the comoving frame
$u^x=u^y=u^z=0$ for hypersurfaces $u^\psi=0$, one obtains $u^t =
\sqrt{g^{tt}}=\sqrt{{\Lambda_0 \over \Lambda(t)}}=\left({t\over
t_0}\right)$.

The normalized solutions for the time-dependent modes of the the
inflaton field and the photons, are
\begin{eqnarray}
&& \bar\Psi_k(t) = \left(\frac{t}{t_0}\right)^{\frac{(1+6p)}{2}}
\Psi_k(t),\qquad \bar\Psi_{k}(t)= \frac{i}{2}
\sqrt{\frac{\pi}{m_0\,p}} \,\,{\cal H}^{(2)}_{\frac{3}{2}
\sqrt{1+\frac{1}{9}\left(\frac{1}{p^2}-4m^2_0\right)}}\left[k\left(\frac{t_0}{t}\right)^p\right], \\
&& \bar\chi_k(t) = \left(\frac{t}{t_0}\right)^{3/2} \chi_k(t),
\qquad \bar\chi_{k}(t) = \frac{i}{2} \sqrt{\frac{\pi}{p}}
\,\,{\cal H}^{(2)}_{\frac{\sqrt{21}}{2}}
\left[k\left(\frac{t_0}{t}\right)^p\right].
\end{eqnarray}
The squared magnetic field fluctuations on the physical and
comoving frames, are
\begin{eqnarray}
&& \left.\left< B^2_{(phys)}\right>\right|_{IR} \simeq
\frac{K}{p^2}\,\,\left(\frac{\Lambda_0}{3}\right)^2 \,
\epsilon^{5-\sqrt{21}}
\left(21\,p^2-1\right)^{\frac{5-\sqrt{21}}{2}}
\left(\frac{t}{t_0}\right)^{-2p}, \\
&& \left.\left< B^2_{(com)}\right>\right|_{IR} \simeq
\frac{K}{p^2}\,\,\left(\frac{\Lambda_0}{3}\right)^2 \,
\epsilon^{5-\sqrt{21}}
\left(21\,p^2-1\right)^{\frac{5-\sqrt{21}}{2}}
\left(\frac{t}{t_0}\right)^{2p},
\end{eqnarray}
where $K$ is a constant of integration. The power spectrums of
$\left<\varphi^2\right>$ and $\left<B^2\right>$, go as
\begin{eqnarray}
&& {\cal P}^{\left<\varphi^2\right>}(k) \sim k^{3\left[1-\sqrt{1+\frac{1}{9}\left(\frac{1}{p^2}
-4m^2_0\right)}\right]}, \label{s3} \\
&& {\cal P}^{\left<B^2\right>}(k) \sim k^{5-\sqrt{21}} \sim
k^{0.42}. \label{s4}
\end{eqnarray}
Note that ${\cal P}^{\left<\varphi^2\right>}(k)$ becomes nearly
scale invariant for $m_0\simeq {1\over 2p} \ll 1$. Furthermore,
the parameter $m_0$ is related to the power of expansion $p$ of
the universe. This result agrees with whole that one expects in
the sense that $m_0$ will be very small for a very accelerated
universe (i.e., for $p \gg 1$).\\

 \section{Final Comments}

In this work we have shown how large-scale magnetic fields with
sufficiently large amplitude can be generated in the early
universe from a 5D vacuum state. We have explored two examples,
which are relevant for cosmology. The first one is the well known
de Sitter expansion. The second describes an universe governed by
a decaying cosmological parameter. In both cases we obtained the
same power for the spectrum of
$\left.\left<B^2\right>\right|_{IR}$, because the origin of this
power is geometrical and depends on the components of the Ricci
tensor on the effective 4D hypersurface. In the examples here
worked these components are the same: $R^i_{\,\,i}=\Lambda_0$. An
important result here obtained is that the power of the spectrums
for $\left.\left<B^2\right>\right|_{IR}$ we found is positive. It
suggests that more intense magnetic fields should be on smaller
scales, which is in agreement with observation. Furthermore, in
both cases $\left.\left<B^2_{(phys)}\right>\right|_{IR} \sim
a^{-2}$ (and not as $a^{-4}$), due to the superadiabatic
amplification of the modes produced during inflation. Notice that
the results obtained in this paper depends of the gauge we choose.
A gauge invariant formalism will be studied in a future work. It
is important to notice that the theory we have worked is not
conformally invariant, but in a natural manner. The origin of this
rupture is in the gravitational contribution (through $g_{BC}$) of
the operator $Q_{BC} = F_{BC} + \gamma \,\,g_{BC}
\left(A^D_{;D}\right)$.

\begin{acknowledgments}
 The authors acknowledge CONICET and UNMdP (Argentina) for financial
support.
\end{acknowledgments}

\end{document}